# Semantic Technology-Assisted Review (STAR)
# Document analysis and monitoring using random vectors

**Jean-François Delpech**
1515 N. Colonial Ct
Arlington, VA 22209-1439
United States
`jfdelpech@gmail.com`

The review and analysis of large collections of documents and the periodic monitoring of new additions thereto has greatly benefited from new developments in computer software. This paper demonstrates how using random vectors to construct a low-dimensional Euclidean space embedding words and documents enables fast and accurate computation of semantic similarities between them. With this technique of Semantic Technology-Assisted Review (STAR), documents can be selected, compared, classified, summarized and evaluated very quickly with minimal expert involvement and high-quality results.

## 1. Introduction

In a recent review article, M. R. Grossman and G. V. Cormack [1] give an extensive discussion of various approaches to technology-assisted review, as well as a very detailed bibliography. They define "Technology-assisted review (TAR) [as] the process of using computer software to categorize each document in a collection as responsive or not, or to prioritize the documents from most to least likely to be responsive, based on a human's review and coding of a small subset of the documents in the collection." Various methods of machine learning have been proposed where a small, human-coded subset of representative documents forms a starting point for machine classification and categorization of the whole collection. The results are presumably more consistent and more reliable than manual review, which is error-prone and not practical for collections of millions of items.

If consideration is limited to textual documents, as is the case here, the starting point is the fact that any document $\mathcal{D}_j$ containing $t$ distinct words can be represented as a vector in an orthonormal vector space where each dimension represents a word $w_{k,j}$ occurring $n_{k,j}$ times:

$$\mathcal{D}_j = (n_{1,j}w_{1,j}, n_{2,j}w_{2,j}, \ldots, n_{t,j}w_{t,j}) \qquad (1)$$

This has been well understood since the pioneering work of Salton [2, 3].

With $M$ documents and $N$ distinct words, the corpus of documents to be reviewed is thus represented by a $M \times N$ term-document matrix with $M$ rows such as Equation 1 and $N$ columns; this matrix is very sparse, since a document will usually contain only a very small fraction of all words (very frequent words, such as *the* or *and* are generally ignored because they have no discriminatory value between documents.) This representation is extremely fruitful and forms the basis of numerous information retrieval systems.

Note that the dual $N \times M$ representation where words are expressed in terms of documents is in principle equivalent but seldom used because it presents a number of practical difficulties.

A first approach relies on keywords for the selection of documents; the SMART method initially developed by G. Salton [2, 3] involves building a reverse index of the collection of documents (at its simplest, just considering each column of the matrix defined by Equation 1). Although much more sophisticated, this is essentially what Apache Lucene [4] does and it can be very useful, for example when searching for a specific word such as a product or an individual name.

However, in Equation 1, each distinct word is orthogonal to each other by the very definition of the embedding space. This has serious practical consequences: since different individuals or organizations use different words to describe the same thing, there is no "best" keyword or set of keywords to retrieve relevant items. For example, the words *spectrum* and *wavelength* have related meanings, but this is completely ignored by a purely keyword-related software. A robust system should automatically take into account the fact that *counterfeiter* and *authentication*, or *fuel*, *combustion* and *injector* are semantically related: if a document contains the word *counterfeiter* and another the word *authentication*, they cover presumably similar topics even if they don't share any word. Deliberate content masking is also a serious problem in document review and analysis: authors do not necessarily seek clarity; in fact, they often prefer some degree of obfuscation for a number of more or less legitimate reasons. Obviously, in the hands of an expert, a sequence of well-designed Boolean queries can be successful, but when millions of documents are involved it is difficult to be certain that coverage is adequate.

To group together words referring to similar topics so that they are not orthogonal to each other, the initial space of dimensionality equal to the number $M$ of distinct, significant words (usually several hundred thousands dimensions) must be transformed to a space of much lower dimensionality, say a few hundred dimensions. Dimensionality reduction, ie. low-rank approximation of the $M \times N$ term-document matrix, can be achieved by a number of techniques which tend to be slow and cumbersome when $M$ and $N$ are both large.

One of the first such techniques was Latent Semantic Indexing [5], which reduces dimensionality through a singular value decomposition (SVD) of the term-document matrix, retaining only a comparatively small number (typically a few hundreds) of the largest singular values. This method has been and is still successfully used for document indexing and retrieval. It suffers nevertheless from serious limitations:

- SVD is computationally intensive, even though the large term-document matrix is very sparse, as it typically depends at least on the square of $M$ or $N$, whichever is largest;
- There is no really satisfactory way to increment the results as new terms/documents become available.

More recently, a number of related methods have been proposed to achieve dimensionality reduction, such as machine learning, neural networks or predictive coding. These related computational techniques provide different measures of similarity between words and/or documents and some of these methods have been discussed and evaluated in the Grossman and Cormack [1] review article quoted above. A striking improvement in speed was demonstrated a few years ago by Mikolov *et al.* [6, 7] who proposed novel model architectures for computing continuous vector representations of words from very large data sets; as a result, to each word is associated a vector with a few hundred floating-point coordinates and the similarity between two words is given by the scalar product between their associated, normalized vectors.

The present article demonstrates that effective vector representations of words and documents can also be obtained simply and economically by using random vectors: over the last ten years there have been several academic publications on the use of random vectors to reduce dimensionality and create a semantic space [8, 9, 10, 11, 12, 13]. Typically, in this context, to each word is attached a random vector with equal numbers of +1 and -1 coordinates (for example, 20 of each) randomly distributed among a larger number of zero coordinates (typically a few hundreds). New and original techniques and algorithms have been developed to (i) compile document collections such as patents, (ii) create the corresponding semantic space, (iii) quantify and limit the noise resulting from the use of random vectors and (iv) retrieve very efficiently information according to users' needs [14].

## 2. Random vectors

Fundamental to a random vectors approach is the fact that, while obviously one cannot create more than $d$ orthogonal vectors in a space of dimension $d$, one can create an exponentially large number of vectors which are quasi-orthogonal to each other; in other words [9, 14], a set of $\exp(\mathcal{O}(\epsilon^2 d))$ vectors picked at random will with high probability be quasi-orthogonal, i.e. have angles of $90 \pm \epsilon$ with each others. The seed vectors referred to below will be selected from such a set $\mathbb{S}_d$ and any linear combination of seed vectors will thus lie in a space of dimension $d$. Instead of being embedded in the very large orthogonal space where each dimension corresponds to a distinct word (millions of distinct words in a typical corpus), each word and combination of words is embedded in a much smaller, quasi-orthogonal space having typically a few hundred to a few thousand dimensions.

The other essential starting point derives directly from Firth's Law of Natural Language Processing (NLP), stating that "you shall know a word by the company it keeps" [15]. The combination of these two fundamental ideas is quite simple in principle:

   a. To each distinct, significant word in a large set of document is associated a random seed vector in a space of dimensionality such that any random vector is with very high probability almost orthogonal to any other.
   b. To each such word is attached a linear combination of the seed vectors of its co-occurring words present, say, in the same window or in the same sentence. This vector lies in a semantic Euclidean space.
   c. Finally, to each document is associated the semantic vector constructed by combining the semantic vectors of each of its words. Words and documents share the same semantic space.

A word is considered significant if it is neither too rare nor too frequent: as noted above, frequent words (words occurring in a large fraction of the documents, for example more than 10%) have obviously little or no discriminatory value between documents; rare words are often typos and their statistical distribution is not significant [16].

If done carefully the process is very quick. There are numerous advantages to an Euclidean space [14], where distance has a well defined meaning: word disambiguation is simply done by Gram-Schmidt orthogonalization, clustering is easy, etc. The computation of the distance or of the similarity between two items, words or documents, reduces to the evaluation of a scalar product, i.e. to a few hundred or thousand floating-point operations and is thus nearly instantaneous: on a small desktop machine, it takes about 1 s. to compute 600,000 scalar products in a single thread. The distance $\mathcal{D}_{ij}$ is related to the similarity $\sigma_{ij}$ by $\mathcal{D}_{ij} = \sqrt{2(1 - \sigma_{ij})}$ and ranges from $0$ (same words $w_i$ and $w_k$) to $2$ (exactly opposite words; note however that owing to the extreme sparsity of a high-dimensional space, the neighborhood exactly opposite a word is in practice always empty.)

The STAR process being entirely linear, the compilation process can be evenly distributed across an arbitrary number of threads and/or processors and the updating process covers only the words contained in the new documents, at least to a very close first approximation.

For the present evaluation, about 814,000 patent applications have been downloaded from the semi-official USPTO site at `http://patentscur.reedtech.com/` between June 2014 and June 2017; these applications cover a broad range of categories and contain 1,430,000 distinct, semantically significant words.

To give an example, the resulting immediate semantic neighborhood of *authenticate* contains *re-authenticate*, *authentication-ok*, *authentication*, *authentication-result*, *biological-characteristic*, *gnubby*, *who-you-are*, *udid-unique*, *protocol-reauthentication*, *dynamic-password*, *descriptionbypattern*, *what-you-know*, *what-you-have*, *per-work-request*, *once-per-connection*, *two-factor*, *static-password* within 50% similarity. Despite typos, the semantic neighborhood is obviously well characterized.

# 3. Examples

## 3.1. Documents close to a reference document

A small extract of the Grossman and Cormack [1] review paper quoted above was used as a reference text:

> *Technology-assisted review (TAR) is the process of using computer software to categorize each document in a collection as responsive or not, or to prioritize the documents from most to least likely to be responsive, based on a human's review and coding of a small subset of the documents in the collection.*

A reference vector $r$ was built as the resultant of the vectors of all significant words appearing in the text, meaning that each coordinate $r_j$ of $r$ is the weighted sum of the corresponding coordinates $t_j^k$, $r_j = \sum_k n^k w^k t_j^k$, where $t^k$ is the vector associated with the $k$-th term (or word), the $w^k$ are standard *tf-idf* statistical weights and $n^k$ is the word's number of occurrences in the text.

This yields the following list of patents where $\sigma$ is the similarity and "Reference" is the USPTO reference:

| | $\sigma$ | Reference | Assignee | Title |
|---|---|---|---|---|
| | | | Table 1 - Patents closest to the Grossman and Cormack[1] definition of TAR (see above) | |
| 1 | 0.76 | 20160371261 | G. V. Cormack, M. R. Grossman,… | Systems and methods for conducting a highly autonomous technology-assisted… |
| 2 | 0.75 | 20160371262 | G. V. Cormack, M. R. Grossman,… | Systems and methods for a scalable continuous active learning approach to information… |
| 3 | 0.75 | 20170132530 | Recommind, Inc., San Francisco,… | Systems and methods for predictive coding |
| 4 | 0.74 | 20160371364 | G. V. Cormack, M. R. Grossman,… | Systems and methods for conducting and terminating a technology-assisted review |
| 5 | 0.69 | 20140279716 | G. V. Cormack, M. R. Grossman,… | Systems and methods for classifying electronic information using advanced active… |
| 6 | 0.69 | 20150324451 | G. V. Cormack, M. R. Grossman,… | Systems and methods for classifying electronic information using advanced active… |
| 7 | 0.69 | 20140280238 | G. V. Cormack, M. R. Grossman,… | Systems and methods for classifying electronic information using advanced active… |
| 8 | 0.68 | 20160371369 | G. V. Cormack, M. R. Grossman,… | Systems and methods for conducting and terminating a technology-assisted review |
| 9 | 0.68 | 20150220519 | Tetsura Motoyama, Los Altos,… | Electronic document retrieval and reporting with review cost and/or time estimation |
| 10 | 0.67 | 20150310068 | Catalyst Repository Systems,… | Reinforcement learning based document coding |
| 11 | 0.66 | 20170116544 | Controldocs.com | Apparatus and method of implementing batch-mode active learning for technology-assisted… |
| 12 | 0.65 | 20170083564 | FTI Inc., Annapolis, US | Computer-implemented system and method for assigning document classifications |
| 13 | 0.65 | 20170116519 | Controldocs.com | Apparatus and method of implementing enhanced batch-mode active learning for… |
| 14 | 0.64 | 20160364299 | Open Text S.A., Luxembourg,… | Systems and methods for content server make disk image operation |
| 15 | 0.64 | 20150178384 | BANK OF AMERICA CORPORATION,… | Targeted document assignments in an electronic discovery system |
| 16 | 0.64 | 20160371260 | G. V. Cormack, M. R. Grossman,… | Systems and methods for conducting and terminating a technology-assisted review |
| 17 | 0.64 | 20160196296 | kCura Chicago, US | Methods and apparatus for deleting a plurality of documents associated with… |
| 18 | 0.63 | 20140317147 | Jianqing Wu, Beltsville,… | Method for improving document review performance |

A list of corporations active in this domain with patent applications submitted in the three-year period covered can easily be obtained by adding the normalized vectors associated with the patents from the top 10 assignees of Table 1 and listing the assignees for the patents closest to this vector (Table 2; note that the list has been substantially condensed for practical reasons.)

| Table 2 - Corporations selected using as a query the reference text of table 1 |
|---|
| For brevity, only the top four patents have been retained for each assignee and the list has been cut off at a similarity of 0.8 |
| **FTI Inc., Annapolis, US** |
| US20170083564A1 (0.92) Computer-implemented system and method for assigning document classifications |
| US20160342572A1 (0.89) Computer-implemented system and method for identifying and visualizing relevant data |
| US20160342590A1 (0.87) Computer-implemented system and method for sorting, filtering, and displaying documents |
| US20140250087A1 (0.86) Computer-implemented system and method for identifying relevant documents for display |
| **G. V. Cormack, M. R. Grossman, Waterloo, CA** |
| US20140280238A1 (0.90) Systems and methods for classifying electronic information using advanced active learning techniques |
| US20160371262A1 (0.89) Systems and methods for a scalable continuous active learning approach to information classification |
| US20160371261A1 (0.89) Systems and methods for conducting a highly autonomous technology-assisted review classification |
| US20150324451A1 (0.87) Systems and methods for classifying electronic information using advanced active learning techniques |
| **Jianqing Wu, Beltsville, US** |
| US20140317147A1 (0.89) Method for improving document review performance |
| US20140358518A1 (0.87) Translation protocol for large discovery projects |
| **BANK OF AMERICA CORPORATION, Charlotte, US** |
| US20150178384A1 (0.87) Targeted document assignments in an electronic discovery system |
| US20150066800A1 (0.87) Turbo batch loading and monitoring of documents for enterprise workflow applications |
| **International Business Machines Corporation, Armonk, US** |
| US20150142816A1 (0.85) Managing searches for information associated with a message |
| US20150347429A1 (0.84) Managing searches for information associated with a message |
| US20170083600A1 (0.84) Creating data objects to separately store common data included in documents |
| US20170116193A1 (0.82) Creating data objects to separately store common data included in documents |
| **Chao-Chin Chang, Taipei City, TW** |
| US20140195904A1 (0.84) Technical documents capturing and patents analysis system and method |
| US20140192379A1 (0.80) Technical documents capturing and patents analysis system and method |
| **EQUIVIO LTD., Rosh Haayin, IL** |
| US20150098660A1 (0.84) Method for organizing large numbers of documents |
| US20150066938A1 (0.84) System for enhancing expert-based computerized analysis of a set of digital documents and methods useful in conjunction… |
| **THE TORONTO DOMINION BANK, Toronto, CA** |
| US20170010841A1 (0.83) Document output processing |
| US20170010842A1 (0.83) Document output processing |
| **Microsoft Technology Licensing, Redmond, US** |
| US20160371258A1 (0.83) Systems and methods for creating unified document lists |
| US20160321250A1 (0.81) Dynamic content suggestion in sparse traffic environment |
| **PatentRatings, Irvine, US** |
| US20160004768A1 (0.83) Method and system for probabilistically quantifying and visualizing relevance between two or more citationally… |
| US20150046420A1 (0.80) Method and system for probabilistically quantifying and visualizing relevance between two or more citationally… |
| **Controldocs.com** |
| US20170116519A1 (0.81) Apparatus and method of implementing enhanced batch-mode active learning for technology-assisted review of documents |
| US20170116544A1 (0.81) Apparatus and method of implementing batch-mode active learning for technology-assisted review of documents |
| **Google Inc., Mountain View, US** |
| US20150169562A1 (0.81) Associating resources with entities |
| US20150169564A1 (0.81) Supplementing search results with information of interest |

Similarly, a list of documents ranked by decreasing similarity to a reference document can be created (the reference here was one of the Controlddocs.com patent applications selected by the short query above):

| | Table 3 - Neighbors of *US 2017/0116519 A1* patent application | | |
|---|---|---|---|
| σ | Reference | Assignee | Title |
| 1  1.00 | 20170116519 | Controldocs.com | Apparatus and method of implementing enhanced batch-mode active learning for… |
| 2  0.99 | 20170116544 | Controldocs.com | Apparatus and method of implementing batch-mode active learning for technology-assisted… |
| 3  0.83 | 20160371262 | G. V. Cormack, M. R. Grossman,… | Systems and methods for a scalable continuous active learning approach to information… |
| 4  0.81 | 20160371261 | G. V. Cormack, M. R. Grossman,… | Systems and methods for conducting a highly autonomous technology-assisted… |
| 5  0.79 | 20140280238 | G. V. Cormack, M. R. Grossman,… | Systems and methods for classifying electronic information using advanced active… |
| 6  0.79 | 20150324451 | G. V. Cormack, M. R. Grossman,… | Systems and methods for classifying electronic information using advanced active… |
| 7  0.79 | 20140279716 | G. V. Cormack, M. R. Grossman,… | Systems and methods for classifying electronic information using advanced active… |
| 8  0.79 | 20170060993 | Skytree, Inc., San Jose,… | Creating a training data set based on unlabeled textual data |
| 9  0.77 | 20160371364 | G. V. Cormack, M. R. Grossman,… | Systems and methods for conducting and terminating a technology-assisted review |
| 10 0.75 | 20170083564 | FTI Inc., Annapolis, US | Computer-implemented system and method for assigning document classifications |
| 11 0.75 | 20170011118 | Microsoft Israel Research… | System for enhancing expert-based computerized analysis of a set of digital… |
| 12 0.74 | 20170039194 | EDCO Health Information Soultions,… | System and method for bundling digitized electronic records |
| 13 0.74 | 20170039519 | MAVRO IMAGING, Westampton,… | Method and apparatus for tracking documents |
| 14 0.74 | 20160071070 | MAVRO IMAGING, Westampton,… | Method and apparatus for tracking documents |
| 15 0.74 | 20150310068 | Catalyst Repository Systems,… | Reinforcement learning based document coding |
| 16 0.74 | 20170140030 | Kofax, Inc., Irvine, US | Systems and methods for organizing data sets |
| 17 0.74 | 20150066938 | EQUIVIO LTD., Rosh Haayin,… | System for enhancing expert-based computerized analysis of a set of digital… |
| 18 0.73 | 20160364299 | Open Text S.A., Luxembourg,… | Systems and methods for content server make disk image operation |
| 19 0.73 | 20160371260 | G. V. Cormack, M. R. Grossman,… | Systems and methods for conducting and terminating a technology-assisted review |
| 20 0.73 | 20160371369 | G. V. Cormack, M. R. Grossman,… | Systems and methods for conducting and terminating a technology-assisted review |
| 21 0.72 | 20150169593 | ABBYY InfoPoisk Moscow,… | Creating a preliminary topic structure of a corpus while generating the corpus |
| 22 0.72 | 20160055424 | IBM, Armonk, NY | Intelligent horizon scanning |
| 23 0.72 | 20150254324 | IBM, Armonk, NY | Framework for continuous processing of a set of documents by multiple software… |
| 24 0.72 | 20140214862 | WAL-MART STORES, Bentonville,… | Automated attribute disambiguation with human input |
| 25 0.72 | 20150254323 | IBM, Armonk, NY | Framework for continuous processing of a set of documents by multiple software… |
| 26 0.71 | 20160239559 | UBIC, Tokyo, JP | Document classification system, document classification method, and document… |
| 27 0.71 | 20160048587 | MSC INTELLECTUAL PROPERTIES… | System and method for real-time dynamic measurement of best-estimate quality… |

### *3.2. Word usage comparison between documents*

As already mentioned (Section 1), the frequent practice of evaluating the similarity between documents by considering only words which occur in both while ignoring semantic similarities can be very misleading because of the vocabulary problem.

When using dimensionality reduction techniques such as STAR, in contrast, co-occurrences are ignored and only semantic proximity is taken into account; documents may be semantically close despite having only a few words in common (or even none): it is enough that their constituent words be semantically close. For example, the two words *barcode* and *ocr-enabled* may rarely occur in the same document but they have a similarity of 0.805 and will tend to draw together documents containing only one of them.

STAR first computes the reference's vector in the semantic space and then explores the neighborhood of this vector. It is thus particularly well suited to full text searching, such as finding documents closest to a reference document which may be a patent, a technical description or a set of reference documents acting as a filter.

In a first example (Table 4) the two patents applications are from the same assignee and share most of their significant words: STAR's scalar is 0.845 but a scalar based only on word co-occurrences is only 0.489. As can be seen, words which are not shared between the two patents (on a white background) are nevertheless semantically very close to the other patent. For example, the word *rankings* occurs 45 times in patent #2 but not at all in patent #1 while nevertheless contributing substantially to the overall similarity, since its similarity to patent #1 is fairly high, at 0.370

Table 4 - Word usage in two patent applications
Comparison between two patent applications from the same assignee
STAR similarity between documents: 0.845
Words on grey background are present in both patents (left AND right), others only in one patent (left OR right)
$\sigma$ is the vector similarity of the word to the *other* patent and $N_{oq}$ is the number of times the word appears in text

| Words in US 2016/0371364 A1 *Systems and methods for conducting and terminating a technology-assisted review* (G. V. Cormack, M. R. Grossman, Waterloo, CA) | | | Words in US 2015/0324451 A1 *Systems and methods for classifying electronic information using advanced active [...]* (G. V. Cormack, M. R. Grossman, Waterloo, CA) | | |
|---|---|---|---|---|---|
| Word | $N_{oq}$ | $\sigma$ | Word | $N_{oq}$ | $\sigma$ |
| technology-assisted | 16 | 0.679 | documents | 148 | 0.510 |
| cormack | 16 | 0.605 | non-relevant | 15 | 0.476 |
| sigir | 4 | 0.588 | move-to-front | 2 | 0.472 |
| documents | 62 | 0.543 | learning | 49 | 0.464 |
| manheimer | 1 | 0.507 | technology-assisted | 5 | 0.438 |
| review | 30 | 0.466 | rankings | 45 | 0.433 |
| bagdouri | 2 | 0.438 | scores | 62 | 0.419 |
| oard | 2 | 0.438 | ranking | 7 | 0.415 |
| non-relevant | 2 | 0.437 | classifier | 42 | 0.414 |
| glanville | 1 | 0.431 | unsupervised | 4 | 0.411 |
| joho | 1 | 0.425 | learn | 1 | 0.408 |
| classification | 38 | 0.411 | supervised | 7 | 0.407 |
| learning | 18 | 0.407 | search | 7 | 0.396 |
| unsupervised | 1 | 0.404 | relevance | 36 | 0.375 |
| supervised | 2 | 0.390 | classification | 48 | 0.367 |
| grossman | 8 | 0.382 | cormack | 3 | 0.358 |
| classifier | 7 | 0.357 | multi-phased | 3 | 0.356 |
| transductive | 1 | 0.355 | departments | 1 | 0.350 |
| search | 16 | 0.353 | manually | 3 | 0.346 |
| machine-learning | 1 | 0.352 | manual | 3 | 0.342 |
| retrieval | 9 | 0.342 | searches | 4 | 0.339 |
| effort | 13 | 0.340 | effort | 6 | 0.336 |
| lefebvre | 1 | 0.328 | classify | 12 | 0.336 |
| reviewer | 8 | 0.324 | classifying | 10 | 0.331 |
| classify | 5 | 0.320 | training | 16 | 0.330 |
| hockeywere | 1 | 0.319 | millions | 1 | 0.326 |

In a second example (Table 5 next page) the two patents do not share many significant words: a scalar based on word co-occurrences is only 0.101, well under any likely notice by a human operator, while STAR's scalar is still 0.707, owing to the fact that their constituent words are seen to be semantically close (see eg. *authenticity-indicating* in #1 and *authenticate* in #2, or *visually* in #1 and *visible* in #2.) Clearly, in many situations, an expert interested in topics covered by patent #1 would be well advised to also consider patent #2.

<table>
<tr><td colspan="6">Table 5 - Word usage in two patent applications<br>Comparison between two patent applications from different assignees<br>STAR similarity between documents: 0.707<br>Words on grey background are present in both patents (left AND right), others only in one patent (left OR right)<br>$\sigma$ is the vector similarity of the word to the *other* patent and $N_{oq}$ is the number of times the word appears in text</td></tr>
</table>

| Words in US 2014/0369569 A1 *Printed authentication pattern for low resolution reproductions* (Document Security Systems, Inc., Rochester, US) | | | Words in US 2014/0270334 A1 *Covert marking system based on multiple latent characteristics* (LASERLOCK TECHNOLOGIES Washington, US) | | |
|---|---|---|---|---|---|
| Word | $N_{oq}$ | $\sigma$ | Word | $N_{oq}$ | $\sigma$ |
| authenticity-indicating | 3 | 0.491 | latent | 51 | 0.421 |
| ink | 10 | 0.432 | mark | 39 | 0.378 |
| shapes | 19 | 0.414 | un-aided | 7 | 0.351 |
| authentic | 8 | 0.405 | led | 2 | 0.348 |
| regularly | 14 | 0.386 | authentication | 13 | 0.343 |
| printing | 5 | 0.361 | lighting | 36 | 0.331 |
| jet | 2 | 0.358 | themselves | 1 | 0.329 |
| non-authentic | 1 | 0.326 | authenticated | 15 | 0.326 |
| authenticity | 9 | 0.322 | counterfeiters | 3 | 0.320 |
| latent | 12 | 0.313 | eye | 10 | 0.319 |
| visually | 4 | 0.312 | else | 1 | 0.318 |
| inks | 2 | 0.308 | visible | 12 | 0.315 |
| checks | 1 | 0.304 | marks | 14 | 0.312 |
| correspondence | 5 | 0.292 | illumination | 3 | 0.302 |
| analyzes | 1 | 0.286 | market | 4 | 0.301 |
| authentication | 3 | 0.283 | mean | 5 | 0.295 |
| reproduction | 8 | 0.277 | black | 2 | 0.293 |
| scans | 1 | 0.267 | authenticate | 4 | 0.290 |
| rendered | 1 | 0.261 | emits | 1 | 0.288 |
| intaglio | 1 | 0.256 | illuminated | 15 | 0.288 |
| reproductions | 3 | 0.254 | cfl | 1 | 0.286 |
| version | 7 | 0.254 | specialized | 7 | 0.280 |
| carbon-based | 7 | 0.253 | counterfeit | 9 | 0.278 |
| green | 1 | 0.238 | broadband | 3 | 0.277 |
| red | 1 | 0.238 | authentic | 6 | 0.275 |
| contrasting | 4 | 0.237 | authenticity | 1 | 0.274 |

STAR has thus very good recall (fraction of relevant instances that have been retrieved over the total amount of relevant instances), substantially better than standard Boolean search with keywords.

In many cases, optimizing recall is the best choice for the kind of full text search involved in patent exploration and other types of Technology Assisted Review. However, depending on the nature of the exploration, STAR alone may exhibit insufficient precision (fraction of relevant instances among the retrieved instances) but this is easily remedied by post-filtering, using for example the Lucene engine [4]; in a combination of the two approaches, Lucene keywords may also be complemented by their closest semantic neighbors.

### 3.3. Patent clusterization

Being able to compute the distance between two patents makes it trivial to compute the distance matrix of a set of patents and to clusterize them. A hierarchical algorithm has been used to clusterize the 160 Giesecke & Devrient patent applications present in the database, as shown in Table 6 next page.

### 3.4. Document summarization

An extractive summary can be created by associating a vector with each paragraph, computing the similarities between each paragraph and the whole document, and keeping only (for example) the six more significant paragraphs [17]. While not as good as a generative summary, this process is much faster and allows quick overviews. Table 7 (next page) shows a summary of the claims section of patent application US 2017/0140030 A1 (Systems and methods for organizing data sets) assigned to Kofax, Inc.

| Table 6 - Top clusters of Giesecke & Devrient patents |
|---|
| Eight top clusters of the 160 Giesecke & Devrient patent applications from June 2014 to June 2017 |

**Cluster #1 split at 0.9999**
    US 2015/0071441 A1 | Methods and system for secure communication between an rfid tag and a reader
    US 2016/0094341 A1 | Methods and system for secure communication between an rfid tag and a reader

**Cluster #2 split at 0.8652**
    US 2015/0098642 A1 | Systems, methods, and computer-readable media for sheet material processing and verification
    US 2015/0097027 A1 | Systems, methods, and computer-readable media for sheet material processing and verification

**Cluster #3 split at 0.8147**
    US 2014/0294174 A1 | Efficient prime-number check
    US 2014/0286488 A1 | Determining a division remainder and ascertaining prime number candidates for a cryptographic application

**Cluster #4 split at 0.8120**
    US 2015/0179013 A1 | Method and apparatus for processing value documents
    US 2017/0158369 A1 | Method and apparatus for processing a transportation container with valuable articles

**Cluster #5 split at 0.7446**
    US 2014/0338457 A1 | Method and apparatus for checking a value document
    US 2014/0352441 A1 | Method and apparatus for examining a value document

**Cluster #6 split at 0.7326**
    US 2014/0297536 A1 | System and method for processing bank notes
    US 2014/0325044 A1 | System and method for processing bank notes
    US 2014/0348413 A1 | Method and apparatus for the determination of classification parameters for the classification of bank notes

**Cluster #7 split at 0.6776**
    US 2015/0258838 A1 | Optically variable areal pattern
    US 2016/0170219 A1 | Optically variable areal pattern

**Cluster #8 split at 0.6485**
    US 2016/0055358 A1 | Check of a security element furnished with magnetic materials
    US 2014/0367469 A1 | Method and apparatus for checking value documents

| Table 7 - Example of an extractive summary |
|---|
| Ellipses **...** stand for skipped paragraphs |

*Classifying one or more test documents into one of a plurality of categories using the binary decision model, wherein the one or more test documents lack a user-defined category label;*

**...**

*Receiving, via the computer and from the user, a confirmation or a negation of a classification label of the most relevant example of the classified test documents; and* **...**

**... ...**

*3. The method of claim 1, wherein the most relevant example of the classified test documents is the test document having a classification score closest to a boundary between a positive decision and a negative decision concerning the test document belonging to a particular one of the plurality of categories.*

**...**

*5. The method of claim 1, further comprising generating a second binary decision model by training the binary classifier using the plurality of training documents and the confirmation or the negation of the classification label of the most relevant example of the classified test documents.*

**... ... ... ... ... ... ...**

*Generating a new binary decision model by training the binary classifier using the plurality of training documents and the confirmation or the negation of the classification label of the most relevant example of the reclassified test documents.*

**... ... ... ... ... ... ... ... ... ... ... ... ...**

*16. The method of claim 15, further comprising selecting one or more relevant examples from among the plurality of organized documents in the problematic category.*

where the ellipses **...** stand for deleted lower-similarity paragraphs; 6 paragraphs out of 60 have been kept.

### 3.5. Portfolio comparison

Portfolio comparison is also another example of how the distance matrix between two sets of patents can be used. Here, the first set of 160 patents belongs to Giesecke & Devrient and the second (6258 patents) to Fujitsu.

| Table 8 - Batch comparison between Giesecke and Fujitsu patent applications |
|---|
| Comparing the 160 Giesecke patents to the 6258 Fujitsu patent applications present in the database |
| (160 x 6258 matrix) |
| Only the top significant results are shown |
| **Fujitsu patents closest to Giesecke patent US 2016/0217442 A1 \| Method for payment** |
| 0.88 \| US 2015/0052054 A1 \| Purchasing service providing method, purchasing service providing apparatus, and recording medium |
| 0.76 \| US 2016/0335855 A1 \| Information providing system and information providing method |
| 0.73 \| US 2014/0297383 A1 \| Information processing apparatus, price calculation method, and recording medium |
| **Fujitsu patents closest to Giesecke patent US 2015/0317268 A1 \| System and method for evaluating a stream of sensor data for value documents** |
| 0.86 \| US 2015/0089480 A1 \| Device, method of generating performance evaluation program, and recording medium |
| 0.71 \| US 2014/0373038 A1 \| Quality evaluation apparatus, quality evaluation method, communication system, and radio base station [...] |
| **Fujitsu patents closest to Giesecke patent US 2015/0026790 A1 \| Method for computer access control by means of mobile end device** |
| 0.82 \| US 2015/0154388 A1 \| Information processing apparatus and user authentication method |
| 0.81 \| US 2015/0128217 A1 \| Authentication method and authentication program |
| 0.81 \| US 2014/0173714 A1 \| Information processing apparatus, and lock execution method |
| 0.80 \| US 2017/0054717 A1 \| Communication method, communication terminal apparatus, and communication network system |
| 0.79 \| US 2015/0256530 A1 \| Communication terminal and secure log-in method |
| 0.78 \| US 2014/0317692 A1 \| Information processing unit, client terminal device, information processing system, and authentication [...] |
| 0.78 \| US 2014/0380440 A1 \| Authentication information management of associated first and second authentication information for user [...] |
| **Fujitsu patents closest to Giesecke patent US 2017/0106689 A1 \| Security element having a lenticular image** |
| 0.81 \| US 2014/0375869 A1 \| Imaging apparatus and imaging method |
| 0.76 \| US 2015/0316779 A1 \| Optical device |
| 0.75 \| US 2016/0209596 A1 \| Inter-lens adjusting method and photoelectric hybrid substrate |
| 0.71 \| US 2014/0347725 A1 \| Image display device and optical device |
| 0.71 \| US 2015/0261000 A1 \| 3d image displaying object, production method, and production system thereof |
| **Fujitsu patents closest to Giesecke patent US 2015/0286473 A1 \| Method and system for installing an application in a security element** |
| 0.80 \| US 2014/0325501 A1 \| Computer installation method, computer-readable medium storing computer installation program, and computer [...] |
| 0.79 \| US 2014/0298321 A1 \| Installation control method and installation control apparatus |
| 0.73 \| US 2016/0112280 A1 \| Data network management system, data network management apparatus, data processing apparatus, and data [...] |
| **Fujitsu patents closest to Giesecke patent US 2015/0071441 A1 \| Methods and system for secure communication between an rfid tag and a reader** |
| 0.77 \| US 2017/0046543 A1 \| Equipment inspection apparatus and equipment inspection method |
| 0.71 \| US 2015/0220762 A1 \| Information reading system, reading control device, reading control method, and recording medium |

### 3.6. Disambiguation of polysemous terms

As shown in the first column of Table 9, the term *mantle* has at least two very different meanings in the patent database: considering its two closest neighbors, it may refer to a common laboratory equipment, a *heating mantle*, often associated with a *stirrer*, or it may refer to a *mantle cell*, often associated in cancerology with *Burkitt lymphoma*.

Since the STAR process results in a quasi-orthogonal Euclidean space, the Schmidt orthogonalization procedure does remove this kind of ambiguity. Assuming term vectors to be normalized to unity, one needs simply to subtract from the vector |mantle⟩ the collinear component of the vector |burkitt⟩ to eliminate the meaning related to *burkitt*:

$$|\text{mantle}\rangle_{\perp burkitt} = |\text{mantle}\rangle - \langle\text{mantle}|\text{burkitt}\rangle \times |\text{burkitt}\rangle \qquad (2)$$

in bra-ket notation with the following result (Table 9), where the meaning related to *burkitt* is totally eliminated in the second column and the meaning related to *stirrer* is totally eliminated in the third:

Table 9 - Top neighbors of *mantle*, *mantle* ⊥ *burkitt*, *mantle* ⊥ *stirrer* and *stirrer*
Schmidt orthogonalization is used to separate different meanings of a polysemous word
⊥ stands for "orthogonalized with respect to"

| mantle | | mantle ⊥ burkitt | | mantle ⊥ stirrer | | burkitt | |
|---|---|---|---|---|---|---|---|
| mantle | 1.000 | stirrer | 0.571 | dsccl | 0.580 | burkitt | 1.000 |
| stirrer | 0.548 | flask | 0.512 | centrocytic | 0.580 | waldenstrom | 0.960 |
| burkitt | 0.547 | claisen | 0.503 | burkitt | 0.570 | macroglobulinemia | 0.941 |
| vigreux | 0.540 | tmbpf | 0.503 | mantle-cell | 0.562 | immunoblastic | 0.935 |
| immunoblastic | 0.536 | round-bottom | 0.495 | lymphoplasmacytic | 0.560 | mediastinal | 0.914 |
| splenic | 0.532 | mantle | 0.476 | malignany | 0.559 | follicular | 0.903 |
| lymphoplasmacytic | 0.531 | rotavapor | 0.474 | histiocyte-rich | 0.558 | hairy | 0.902 |
| mediastinal | 0.526 | vigreux | 0.469 | enteropathy-type | 0.558 | plasmacytoma | 0.900 |
| waldenstrom | 0.524 | kettle | 0.464 | prolymphocytic | 0.555 | lymphoplasmacytic | 0.899 |
| follicular | 0.521 | four-neck | 0.464 | mediastinal | 0.553 | splenic | 0.897 |
| mantle-cell | 0.521 | hpvp | 0.461 | immunoblastic | 0.553 | immunocytoma | 0.896 |
| extranodal | 0.519 | t-bhp | 0.460 | extra-nodal | 0.551 | prolymphocytic | 0.895 |
| prolymphocytic | 0.513 | multi-neck | 0.456 | lymphoplasmocytic | 0.548 | histiocyte-rich | 0.895 |
| extra-nodal | 0.511 | separatory | 0.456 | hepatosplenic | 0.547 | b-lymphoblastic | 0.878 |
| scfh | 0.510 | three-neck | 0.455 | extranodal | 0.547 | extranodal | 0.876 |
| b-lymphoblastic | 0.510 | stark | 0.452 | splenic | 0.544 | mixed-cellularity | 0.874 |
| malignany | 0.509 | exotherm | 0.451 | eatl | 0.540 | monocytoid | 0.865 |
| flask | 0.508 | sparge | 0.449 | t-lymphoblastic | 0.537 | smzl | 0.863 |
| sparge | 0.508 | dean | 0.448 | b-lymphoblastic | 0.537 | lymphomatoid | 0.860 |
| histiocyte-rich | 0.507 | stirring | 0.443 | histiocyte | 0.535 | nmzl | 0.858 |
| macroglobulinemia | 0.505 | stiffing | 0.443 | nmzl | 0.534 | histiocyte | 0.848 |

### 3.7. Variability and noise

There are several sources of variability and/or noise in any method relying on textual word proximity, whether SVD, machine learning, neural networks, predictive coding or STAR.

a. A fundamental source of variability is due to the randomness of the database; while the co-occurrences of frequent words are fairly stable, this is obviously not the case for rare words occurring from a few times to a few dozen times. If the database had not included biology and medicine, for example, the word *burkitt* would most probably not have shown up as a close neighbor of *mantle*, independently of the number of words in the database (in this case, more than two billions [18].)

b. A second source of variability occurs from differences in what is understood by the word "neighbor". In this work, it was defined as "belonging to the same sentence": the weighted sum of all significant word vectors in a sentence was added to create a sentence vector, which was then added to the vectors of each word in the sentence (this was experimentally found to be a good choice for patent analysis.) However, depending on the result to be achieved, other definitions would be just as acceptable [19, 20]. For example, limiting grouping to a five-word window does favor synonyms over simple neighbors: in this case, *burkitt*, which usually appears in the same sentence as *mantle* but at a distance of several words, would not have been listed as a close neighbor of *mantle*.

c. Some noise arises from the random vector representation itself. In this work, as the embedding space is only quasi-orthonormal, two randomly chosen seed vectors will in general have a small, but non-zero scalar product. As shown previously [14], this adds a zero-centered Gaussian noise to the scalar product of randomly chosen vectors. This noise decreases as the square root of the dimension $d$ of the embedding space and is in general negligible in comparison to the variability associated with other causes. All other approaches relying on word proximity have their own sources of noise; for example, Mikolov *et al.* [6, 7] initialize their computations with random coefficients, their negative sampling method relies on randomly drawing words from the corpus and their technique of "subsampling" is also random-based.

## 4. Conclusions

Although the examples given here are drawn from a patent database, the STAR technique can be applied to any corpus of documents. Cormack and Grossman [21], in their 2014 evaluation of machine-learning protocols for TAR, give a few examples of "requests for production" which can be used "as is" to initiate a review. Using just the words "prepay transactions" as query, for example (Matter 201 in their Table 2), generates a list of patents which would probably not be very relevant in a legal situation, but which center around words such as *debit*, *financial*, *credit*, *payment*, *transactions*, *debited*, *institutions*, *transaction*, *accounts*, *funds*, *credited*, *settlement*, with similarities to the query ranging from 0.78 to 0.49. In any real world situation, the best way to initiate a semantic technology-assisted review would probably be (a) selecting the documents which come up with one or several initial requests (first tier), (b), selecting the second-tier documents closest to the first tier and (c) automatically forming clusters of documents for manual review. In many cases, a reasonable similarity threshold between documents appears to be around 0.7. Once a semantic space has been automatically constructed from the corpus, the process illustrated by Tables 1, 2, 6 and 8 above is very quick and requires very little operator input. This approach has some similarity to the CAL protocol advocated by Cormack and Grossman [21]; a test of it in a realistic, legal environment would be of interest.

The STAR technique has also several obvious advantages for intellectual property rights assessments; for example, in the case of patents, once a suitable database has been collected and a semantic vector space has been constructed, STAR is well suited to examine issues such as patentability by comparison to prior art as well as freedom to operate by detecting potential infringements. With STAR, performing a patent or technology watch simply involves setting-up a filter and periodically checking for new information, as was done above in Section 3.5; this can be personalized with minimal effort for an arbitrary number of clients

All of these examples involve comparing a document or a set of documents to documents present in the database, either covering a definite time period (e.g. last week or last month, typically several thousand US patent applications) or covering the whole database (in actual production, several million patents.)

In addition to patents, the database may include any other kind of textual documents, such as technical publications, descriptions of technologies under development "in house", patent projects, or even highly speculative ideas. With STAR, even a query based on a short (e.g. one page or even one sentence or one phrase) description should in most situations be enough to generate a reasonably short, but quite relevant, ranked list of the documents closest to the query.